\documentclass[pre,twocolumn,showpacs,superscriptaddress]{revtex4}

\usepackage[german,english]{babel}
\usepackage{amssymb}
\usepackage{amsfonts}
\usepackage{amsmath}

\begin{document}

\title{Entropy loss in long-distance DNA looping}

\author{Andreas Hanke}
\affiliation{Institut f{\"u}r Theoretische Physik, Universit{\"a}t
Stuttgart, Pfaffenwaldring 57, D-70550 Stuttgart, Germany}
\affiliation{Department of Physics $-$ Theoretical Physics, 
University of Oxford, 1 Keble Road, Oxford OX1 3NP, United Kingdom}
\author{Ralf Metzler}
\email{metz@nordita.dk}
\affiliation{NORDITA, Blegdamsvej 17, DK-2100 Copenhagen {\O}, Denmark}
\affiliation{Department of Physics, Massachusetts Institute of Technology,
77 Massachusetts Avenue, Cambridge, Massachusetts 02139, USA}

\date{\today}

\begin{abstract}
The entropy loss due to the formation of one or multiple loops in circular
and linear DNA chains is calculated from a scaling approach in the limit
of long chain segments. The analytical results allow to obtain a fast
estimate for the entropy loss for a given configuration. Numerical
values obtained for some examples suggest that the entropy loss encountered
in loop closure in typical genetic switches
may become a relevant factor which has to be overcome by
the released bond energy between the looping contact sites.
\end{abstract}

\pacs{87.15.Ya,02.10.Kn,02.50.-r,87.15.Aa}

\maketitle

\section{Introduction}

\begin{figure}
\unitlength=1cm
\begin{picture}(6,4)
\put(0.4,4.4){\includegraphics{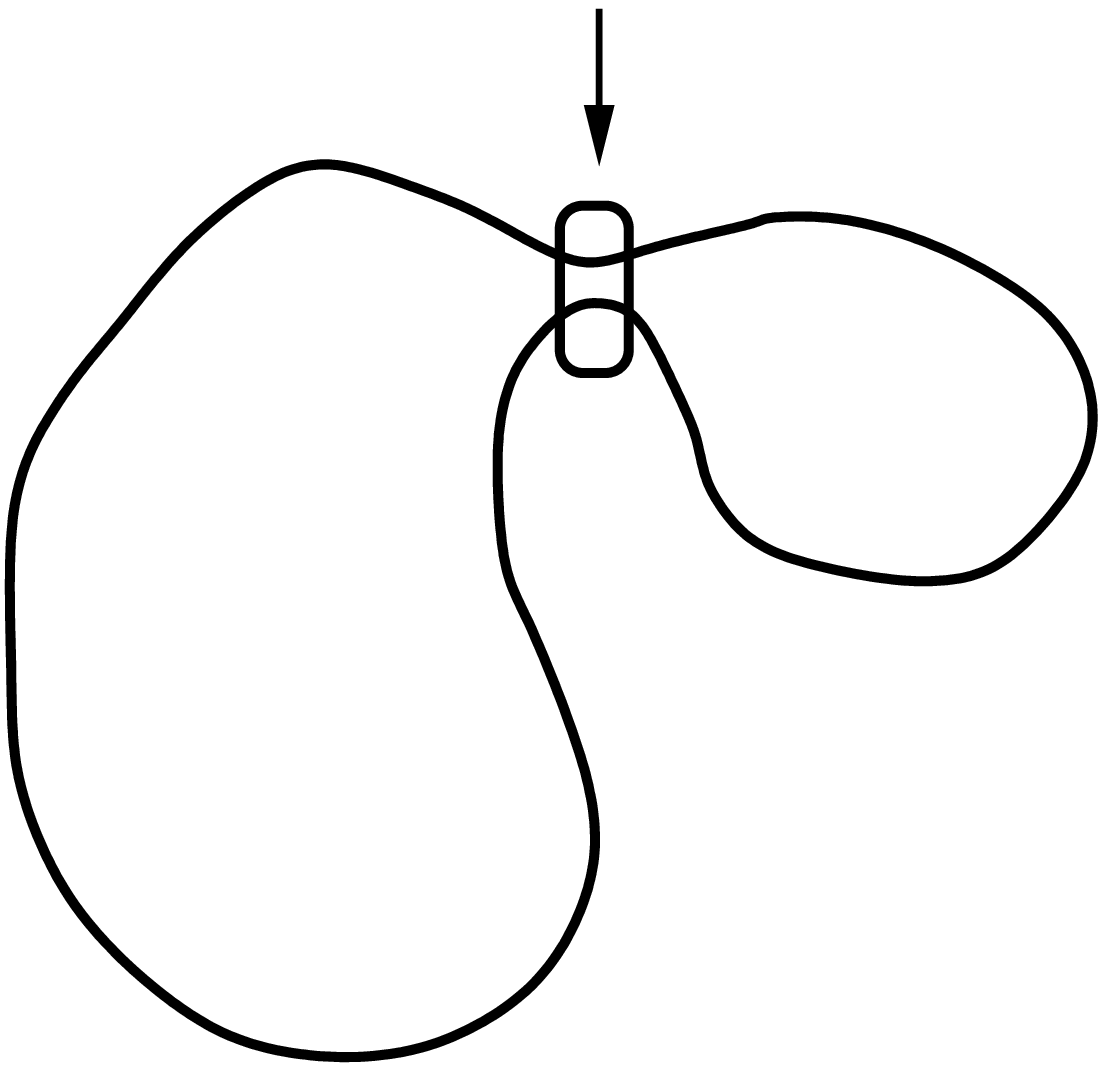}}
\put(4.2,3.8){\includegraphics{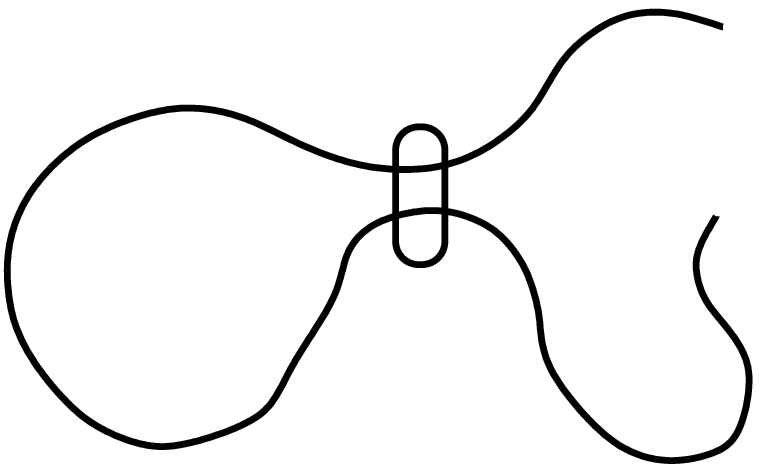}}
\put(-0.4,1.9){$L-\ell$}
\put(0,0.2){$\ell$}
\put(4.3,2){$\ell$}
\put(3.9,0.9){$\ell_1$}
\put(6,0.7){$\ell_2$}
\end{picture}
\caption{DNA looping in a circular (left) and linear DNA (right). The rounded
boxes indicate the chemical bonds established between the transcription factors
through looping at specific contact sites on the DNA double-helix which are
fairly distant from one another in terms of the arc length along the DNA.
A telomere loop corresponds to the right configuration with vanishing $\ell_1$
[or $\ell_2$].
\label{fig1}}
\end{figure}

Gene expression in all organisms comprises the transcription of a certain
gene on the DNA into messenger RNA (mRNA) through RNA polymerase starting
from the promoter site, and its subsequent translation into a protein. The
initiation of the transcription at a specific gene underlies a subtle
cooperative scheme of transcription factors, which in turn is determined
by a given set of boundary conditions such as the concentration of the
transcription factors. Transcription factors often act cooperatively, and
they are known to interact with each other over distances of
several thousand base pairs (bp). This interaction is effected through
DNA looping, compare Fig.~\ref{fig1} \cite{alberts,genetics,ptashne,revet}.

A typical example for DNA looping is found in the genetic switch which
determines whether the replication of bacteriophage $\lambda$ in E.coli
follows either the lysogenic or the lytic pathway \cite{genetics,ptashne}.
A key component of this $\lambda$ switch is the $\lambda$ repressor which
activates
the expression of a gene that encodes the production of the $\lambda$
repressor itself. $\lambda$ repressor can bind to the three operator
sites O$_{\rm R}$ which overlap the two promoter sites of the switch.
$\lambda$ repressor binds cooperatively as a dimer, and typically under
stable lysogenic conditions two such dimers on O$_{\rm R}$ form a tetramer,
the next higher order of cooperativity, which is the main factor for the
stability of the $\lambda$ switch against noise \cite{gsw}.
However, $\lambda$ repressor
can also bind to the very similar operator O$_{\rm L}$, which is located
roughly 2300 bp away and not part of the $\lambda$ switch. It has been
found that the two $\lambda$ repressor tetramers at O$_{\rm L}$ and
O$_{\rm R}$ synergistically form an octamer through DNA looping. This
higher-ordered oligomerisation enhances the performance of the switch
considerably \cite{ptashne,revet,bell,xu,semsey,amouyal}. The specific
binding along the tetramer-tetramer interface has recently been revealed
through crystallographic structure determination \cite{bell}. Similar
realisations of DNA looping also occur in linear DNA, naturally in the form of
telomeres or in vitro in engineered DNA, compare Fig.~\ref{fig1}
\cite{zaman,griffith}. Multiple looping
in large DNA molecules around a locus can be observed in vivo and can be
induced in vitro by introducing of specific binding zones on the DNA, which
leads to a considerable reduction of the gyration radius of the molecule
such that it can be more easily transferred into (e.g., mammalian) cells
\cite{montigny}.

DNA looping often involves large loop sizes of several thousand bp.
Therefore, the formation of these loops causes a non-negligible entropy
loss which has to be overcome by the binding energy released at the
bond formation on loop closure. In the present study, we quantify this
entropy loss for such long DNA loops, taking into account self-avoiding
effects due to both the monomer-monomer interaction within the loop and
the additional effects due to the higher order contact points (vertices)
at the loop closure site. The resulting numbers for typical systems
suggest that the entropy loss is a relevant factor in the formation
of DNA loops, and it gives a lower bound for the bond-forming energy
required to stabilise the loop.

Entropy loss due to loop formation was studied for the case of disconnected
loops by Schellmann and Flory \cite{schellmann}. In their seminal paper,
Poland and Scheraga \cite{scheraga} considered coupled Gaussian loops.
To our knowledge the full effect of self-avoidance in the DNA looping network
has not been considered before. Hereby, the contributions of non-trivial
vertices turns out to be a relevant factor, and for multiple looping with a
common locus actually become the dominating contribution.
The analytical results presented here are
derived from a scaling approach for general polymer networks and provide the
advantage that on their basis estimates for the entropy loss
in a given DNA system can be computed in a straightforward manner.
It should also be noted that the additional vertex effects studied herein
may be crucial in the analytical treatment of the DNA looping {\em
dynamics\/}, as the higher-order self-interaction
at such vertices poses an additional barrier in the loop closure process
\cite{merlitz}. Our results for long DNA with large loops complement the
investigations of the bending and twisting energies in small DNA plasmids
\cite{swigon}. In the case of intermediate sized DNA segments, both approaches
may be combined.

In what follows, we calculate the scaling results for the entropy loss on
looping for the three different cases: (i) looping in a circular DNA, (ii)
looping in a linear DNA, and (iii) multiple looping in a circular DNA. In
the appendix, the general expressions for calculating the system entropy
of an arbitrary polymer network are compiled so that the entropy loss for
different configurations can be calculated according to the general 
procedure developed below.

\section{Looping in a circular DNA chain}

As stated before, we consider the limit in which each segment of the
looped DNA, e.g., both subloops created in the circular DNA upon
looping, are long in comparison to the persistence length $\ell_p$
of the double-stranded DNA chain \cite{rem}. In this long chain limit,
we can neglect energetic effects due to bending or twisting, such that
we treat the DNA as a flexible self-avoiding polymer. Therefore,
we can employ results for the configuration number of a general polymer
network, which we briefly review in the appendix.

Before looping, the free energy of the circular DNA of total 
length $L$ is given by
\begin{equation}
\label{free}
F_{\rm circ}=H_0-TS_{\rm circ} \, ,
\end{equation}
where $H_0$ combines all binding enthalpies in the macromolecule
and the entropy $S_{\rm circ}=k_B\ln\omega_{\rm circ}$ is 
determined by the number of configurations \cite{degennes} 
(see the appendix)
\begin{equation} \label{circ}
\omega_{\rm circ} = A_{\rm circ} \, \mu^L L^{- 3\nu}
\end{equation}
of a simply connected ring polymer of length $L$ in units of 
the monomer length. The latter can be estimated by
the persistence length $\ell_p$ of the polymer
(about 500~{\AA} for double-stranded DNA corresponding to
100 bp \cite{marko}). In equation (\ref{circ}),
$A_{\rm circ}$ is a non-universal amplitude, $\mu$
is the support dependent connectivity constant,
and $\nu \simeq 0.588$ \cite{GZJ} is the Flory exponent.
Thus, $S_{\rm circ}$ has the form
\begin{equation}
S_{\rm circ} = k_B\left(\ln A_{\rm circ} + L\ln\mu - 3\nu\ln L\right)\,.
\end{equation}

On looping, as sketched in Fig.~\ref{fig1} to the left, the 
circular DNA is divided into two subloops of lengths $\ell$ and 
$L-\ell$ by creation of a vertex at which four legs of the chain are 
bound together. For a 
self-avoiding chain, the number of configurations of the resulting 
``figure-eight'' shape \cite{slili2d} is not simply the product of 
the configuration numbers of the two created loops, but has the 
more complicated form \cite{duplantier,binder,schaefer}
(see the appendix)
\begin{equation}
\label{fig8}
\omega_8 = A_8 \, \mu^L(L-\ell)^{-6\nu+\sigma_4}\,{\cal Y}_8
\left(\frac{\ell}{L-\ell}\right) \, .
\end{equation}
In this expression, $A_8$ is a non-universal amplitude,
${\cal Y}_8$ is an universal 
scaling function, and $\sigma_4 \simeq - 0.48$
is an universal exponent associated 
with the vertex with four outgoing legs. Note that in the Gaussian chain
limit, the exponents $\sigma_N$ vanish; as we are going to show, the
inclusion of the additional effects due to the higher order vertex formation
reflected by nonzero values for $\sigma_N$ are non-negligible.
Given the entropy $S_8=k_B\ln\omega_8$ 
of the figure eight configuration, the 
entropy loss suffered from creating this configuration out of
the original circular DNA amounts to
$\left|S_8-S_{\rm circ}\right|$. 
To proceed, we now evaluate the
scaling function ${\cal Y}_8(x)$ in some special cases, 
and calculate typical numbers for the required entropy loss 
compensation. Two limiting cases can be distinguished:

\vspace*{0.2cm}

{\bf (1.)} If one of the loop sizes is much smaller than the other ($\ell\ll
L-\ell$, say), the big loop of size $L-\ell$ will essentially behave like a
free circular chain so that its contribution to $\omega_8$ will scale like a
regular ring polymer, i.e., like $(L-\ell)^{-3\nu}$. Consequently,
we find the behaviour 
${\cal Y}_8(x) = a \, x^{-3\nu+\sigma_4}$ for $x \ll 1$, where
$a$ is an universal amplitude, and therefore
\cite{kafri,slili2d}
\begin{equation}
\omega_8 = A_8 \, a \, \mu^L(L-\ell)^{-3\nu}\ell^{-3\nu+\sigma_4} \, .
\end{equation}
In this case, 
the free energy difference between the initial circular 
and the looped states becomes
\begin{equation}
\Delta F=\Delta H_{\rm bond}-T(S_8-S_{\rm circ}) \,,
\end{equation}
where $\Delta H_{\rm bond}$ is the binding enthalpy at the loop closure site.
The formation of the looping bond has to release a higher enthalpy than what
is lost in entropy, i.e., $\Delta H_{\rm bond}<T(S_8-S_{\rm circ})$.
Collecting the different expressions, we thus find the condition
\begin{equation}
\label{bond1}
\Delta H_{\rm bond}<k_BT\left[
\ln \frac{A_8 \, a}{A_{\rm circ}} + 
3\nu\ln\frac{L}{\ell(L-\ell)}+\sigma_4\ln\ell
\right] \, .
\end{equation}
In this expression (and in similar expressions below), the first
term in the square brackets is non-universal and depends on details
of the model \cite{remark}, whereas the remaining
contributions are universal (apart from the fact that $L$ is 
measured in units of the non-universal monomer length).

To get an estimate for the magnitude of the entropy loss, consider the case
of the $\lambda$ repressor loop in E.coli. With the size of the entire DNA
of approximately $3.5\times 10^3$~kbp and the looping branch of about 2.3~kbp,
the two loops correspond to $3.5 \times 10^4$ and 23 monomers,
respectively (each monomer corresponds to a persistence length 
$\ell_p$ of 100 bp, see above).
Neglecting the non-universal first term in
brackets in expression (\ref{bond1}) \cite{remark}, 
these numbers produce
\begin{equation}
\label{bonden}
\Delta H_{\rm bond}<-7.0~k_BT=-17.5~{\rm kJ}/{\rm mol}=-4.2~{\rm kcal}
/{\rm mol};
\end{equation}
here and in the following examples we choose $T=300^{\circ}{\rm K}$
and make use of the gas constant, 
$R=8.31~{\rm J}{\rm K}^{-1}{\rm mol}^{-1}$, and the
conversion factor $1~{\rm cal}=4.2~{\rm J}$ \cite{abramowitz}.
Expression (\ref{bonden}) gives a considerable minimal value
for the required bond energy between the two looping sites. 
For comparison, the typical free energy for base pair formation 
in DNA is 8 kcal/mol for AT pairs and 13 kcal/mol for GC pairs 
\cite{breslauer}. 
Thus, even for the relatively small loop of 23 monomers, the required
enthalpy release is non-negligible. Note that the relative contribution
stemming from the $\sigma_4$ term in equation (\ref{bond1}) amounts to
about 20\% of the required enthalpy.

\vspace*{0.2cm}

{\bf (2.)} If the two created loops are of comparable size, i.e., 
$x = \ell / (L-\ell) \approx 1$, the corresponding value
of the scaling function ${\cal Y}_8(x)$ is a finite number.
For example, for $\ell=L/2$ one finds
\begin{equation} \label{half}
\Delta H_{\rm bond}<k_BT
\left[\ln \frac{A_8 \, {\cal Y}_8(1)}{A_{\rm circ}} +
\sigma_4\ln\frac{L}{2} - 3\nu\ln\frac{L}{4}
\right] \, .
\end{equation}
In a modified DNA with two loops of 2.3~kbp each, one finds a bond 
enthalpy requirement of
\begin{equation}
\Delta H_{\rm bond}<-5.8~k_BT=-14.5~{\rm kJ}/{\rm mol}=-3.4~{\rm kcal}/
{\rm mol} \, ,
\end{equation}
where we again neglect the non-universal first term in the square brackets 
\cite{remark}.
If both loops are of size $2 \times 10^3$\,kbp each, the required bond
enthalpy would increase to $\Delta H_{\rm bond}<-12.5~{\rm kcal}/{\rm mol}$.

\section{Looping in a linear DNA chain}

A linear chain of length $L$ can assume
\begin{equation}
\omega_{\rm lin} = A_{\rm lin} \, \mu^LL^{\gamma -1}
\end{equation}
distinct configurations, where $A_{\rm lin}$ is a non-universal 
amplitude and $\gamma \simeq 1.16$ is an universal exponent 
\cite{GZJ,caracciolo}. 
If looping occurs and produces the
A-shape in Fig.~\ref{fig1} to the right, the configuration number 
is modified to
\begin{equation}
\omega_{\rm A} = 
A_{\rm A} \, \mu^L (L-\ell)^{\gamma-1-3\nu+\sigma_4} \, {\cal Y}_{\rm A}
\left(\frac{\ell}{L-\ell},\frac{\ell_1}{\ell_2}\right) \, ,
\end{equation}
where $\ell$ is the size of the loop and $\ell_1$, $\ell_2$
are the sizes of the two loose end-segments, respectively.

We distinguish four different cases belonging to two groups: the 
configuration with $\ell_1\approx\ell_2$, and the
telomere configuration for which $\ell_1=0$ (or $\ell_2=0$).

\vspace*{0.2cm}

{\bf (1.)} If $\ell_1=\ell_2$, we find
\begin{equation}
\omega_{\rm A} = 
A_{\rm A} \, \mu^L(L-\ell)^{\gamma-1-3\nu+\sigma_4}\,{\cal W}_{\rm A}
\left(\frac{\ell}{L-\ell}\right) \, ,
\end{equation}
where ${\cal W}_{\rm A}(x) = {\cal Y}_{\rm A}(x,1)$.
If furthermore $\ell\ll L-\ell$, an analogous reasoning as in 
case II.(1.) leads to
\begin{equation}
\omega_{\rm A} = A_{\rm A} \, b \,
\mu^L(L-\ell)^{\gamma-1}\ell^{-3\nu+\sigma_4} \, ,
\end{equation}
where $b$ is an universal number. The fact that $\ell$ carries 
the same exponent as in the above case, number
(1.), is due to the local effect of self-interaction for the small 
loop; in both cases, the small loop is connected to a 4-vertex.

For the binding enthalpy we obtain the condition
\begin{equation}
\Delta H_{\rm bond}<k_B T
\bigg[\ln \frac{A_{\rm A} b}{A_{\rm lin}} +
(\gamma-1)\ln\frac{L-\ell}{L} - (3\nu-\sigma_4)\ln\ell
\bigg] \, .
\end{equation}
To obtain a numerical value, consider the $\lambda$ repressor loop of 23
monomers and the E.coli DNA length $3.5 \times 10^4$ monomers, a configuration
which can be obtained by cutting the E.coli DNA. Neglecting the 
(non-universal) first term in the square brackets, we find in this case
\begin{equation}
\Delta H_{\rm bond}<-7.0~k_BT=-17.5~{\rm kJ}/{\rm mol}=
-4.2~{\rm kcal}/{\rm mol},
\end{equation}
where the exact numerical value is slightly smaller than in equation
(\ref{bonden}).

\vspace*{0.2cm}

{\bf (2.)} Conversely, if $\ell=\ell_1=\ell_2$, the simpler expression
\begin{equation}
\omega_{\rm A} = A_{\rm A} \, 
{\cal W}_{\rm A}(\textstyle{\frac{1}{2}}) \displaystyle \,
\mu^L \left(\frac{2 L}{3}\right)^{\gamma-1-3\nu+\sigma_4}
\end{equation}
emanates, and the binding enthalpy has to fulfil
\begin{eqnarray} \label{bonden1}
\Delta H_{\rm bond} & < & k_BT\bigg[
\ln \frac{A_{\rm A} \, 
{\cal W}_{\rm A}(\frac{1}{2})}{A_{\rm lin}} \nonumber \\
& & + (\gamma-1)\ln\frac{2}{3}-(3\nu-\sigma_4)\ln \frac{2 L}{3}\bigg] \, .
\end{eqnarray}
Taking 23 monomers for each segment and neglecting the
(non-universal) first term in the square brackets yields the condition
\begin{equation}
\Delta H_{\rm bond}<-8.6~k_BT=-21.6~{\rm kJ}/{\rm mol}=
-5.1~{\rm kcal}/{\rm mol}
\end{equation}
for the binding energy.
If the segments are larger by a factor of 100, this value gets modified to
$\Delta H_{\rm bond}<-11.3~{\rm kcal}/{\rm mol}$. 

\vspace*{0.2cm}

{\bf (3.)} The next two cases belong to the telomere configuration
corresponding to Fig.~\ref{fig1} to the right with $\ell_1 = 0$ and
$\ell_2 = L - \ell$. 
This case involves a 3-vertex instead of a 4-vertex, and
has only one loose end-segment. The number of configurations 
the telomere configuration can assume is
\begin{equation}
\omega_{\rm telo} = A_{\rm telo} \, 
\mu^L(L-\ell)^{-3\nu+\sigma_3+\sigma_1}{\cal X}_{\rm
telo}\left(\frac{\ell}{L-\ell}\right) \, ,
\end{equation}
where $\sigma_3 \simeq - 0.18$ and $\sigma_1 = (\gamma-1)/2 \simeq 0.08$
(see the appendix).
Let us first calculate the entropy
loss in the small loop limit $\ell\ll L-\ell$. Here, the linear
chain part should essentially behave like a simple linear chain,
which implies
${\cal X}_{\rm telo}(x) = c \, x^{-3\nu+\sigma_3-\sigma_1}$ 
for $x \ll 1$ and thus
\begin{equation}
\omega_{\rm telo} = A_{\rm telo} \, c \,
\mu^L(L-\ell)^{\gamma-1}\ell^{-3\nu+\sigma_3-\sigma_1} \, ,
\end{equation}
where $c$ is an universal number.

The corresponding condition for the bond enthalpy reads
\begin{eqnarray}
\nonumber
\Delta H_{\rm bond}&<&k_BT\bigg[
\ln \frac{A_{\rm telo} \, c}{A_{\rm lin}} + 
(\gamma-1)\ln\frac{L-\ell}{L}\\
&&-\left(3\nu-\sigma_3+\frac{\gamma-1}{2}\right)\ln\ell\bigg] \, .
\end{eqnarray}
Taking a loop of 2.3~kbp in a chain of length 3500~kbp
and neglecting the (non-universal) first term in the square brackets 
gives 
\begin{equation}
\Delta H_{\rm bond}<-6.3~k_BT=-15.8~{\rm kJ}/{\rm mol}=
-3.8~{\rm kcal}/{\rm mol} \, .
\end{equation}
For comparison, if the loop size is 230~kbp,
this value is increased to 
$\Delta H_{\rm bond}<-9.3~{\rm kcal}/{\rm mol}$.

\vspace*{0.2cm}

{\bf (4.)} If the loop size and the linear chain segment are of equal size,
$\ell=L/2$, the configuration number becomes
\begin{equation}
\omega_{\rm telo} = A_{\rm telo} \, {\cal X}_{\rm telo}(1) \,
\mu^L \left(\frac{L}{2}\right)^{-3\nu+\sigma_3+\sigma_1}
\end{equation}
and we obtain the condition
\begin{eqnarray}
\Delta H_{\rm bond} & < & k_BT\bigg[
\frac{A_{\rm telo} \, {\cal X}_{\rm telo}(1)}{A_{\rm lin}} \nonumber \\
& & - (3\nu-\sigma_3) \ln\frac{L}{2}
- \frac{\gamma-1}{2} \ln(2 L) \bigg] \, .
\end{eqnarray}
Taking a chain length of 460~kbp
and neglecting the (non-universal) first term in the square brackets
we find
\begin{equation}
\Delta H_{\rm bond}<-15.8~k_BT=-39.3~{\rm kJ}/{\rm mol}=
-9.4~{\rm kcal}/{\rm mol}.
\end{equation}

\section{Multiple looping in a circular DNA chain}

\begin{figure}
\unitlength=1cm
\begin{picture}(6,5.2)
\put(-1.1,0.4){\includegraphics{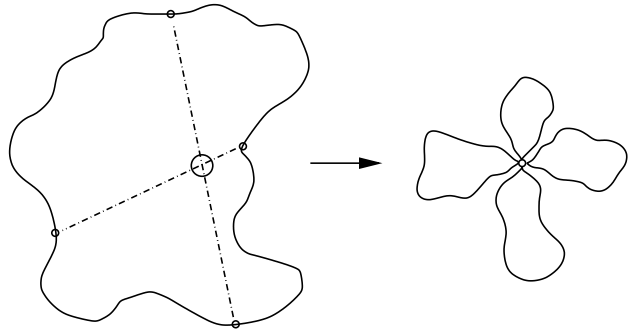}}
\end{picture}
\caption{DNA loop condensation. The circles in the original DNA 
double-helix denote likely contact points. Formation of bonds between these
contacts with one common agglomeration centre, as indicated by the dashed
lines, result in the locus
configuration on the right. Note the reduction in the gyration radius
during this process. A higher-order vertex is created at the locus
point \protect\cite{montigny}.
\label{fig2}}
\end{figure}
Assume that $m$ potential connector points are distributed evenly along a 
circular DNA chain of total length $L$.
If these condense to form a common locus, a number $m$ of loops of
equal size are
created which are held together at this locus, as sketched in Fig.~\ref{fig2}
\cite{montigny}. This creates, in the scaling limit, a high-order vertex
where $2m$ legs are joined. The procedure for the configuration number for
this locus configuration yields
\begin{equation}
\omega_{\rm locus} = A_{\rm locus} \, 
\mu^L \left(\frac{L}{m}\right)^{-3m\nu+\sigma_{2m}} \, ,
\end{equation}
where the universal exponent $\sigma_{2m}$ is associated with a
vertex with $2 m$ outgoing legs (see the appendix).
It should be noted that this result holds true only if the size of the 
locus is much smaller than the sizes of the created loops \cite{slili2d}.

Due to the assumption that all $m$ loops are of the same size,
we immediately arrive at 
\begin{eqnarray}
\nonumber
\Delta H_{\rm bond}&<&k_BT
\bigg[ \ln \frac{A_{\rm locus}}{A_{\rm circ}} +
3 \nu(1-m)\ln L\\
&&+3m\nu\ln m+\sigma_{2m}\ln \frac{L}{m} \bigg] \, .
\end{eqnarray}
The absolute value of $\sigma_{2m}$ increases rapidly with increasing
$m$, and can be determined from Pad{\'e} or Pad{\'e}--Borel analysis
as shown in reference \cite{schaefer}. We list the topological exponents
up to order 8 in the appendix. Taking a circular chain of 3500~kbp
and $m=4$, and neglecting the (non-universal) first 
term in the square brackets, we find that the entropy loss is fairly high
(using $\sigma_8 = -2.4$), 
\begin{equation}
\Delta H_{\rm bond}<-67.4~k_BT=-168~{\rm kJ}/{\rm mol}=
-40.1~{\rm kcal}/{\rm mol} \, .
\end{equation}
In this case, the contribution due to the $\sigma_8$ term is as 
large as 50\% of the total entropy loss.

\section{Conclusions}

We have presented an analytical method to estimate the entropy loss in
different scenarios of DNA looping in the limit of long segments. This
approach takes explicitly the self-avoidance and interacting nature of
the formed loops and other segments into account, and considers the
additional effect of vertex formation, i.e., the effective interaction
between different segments at the point where they are joined.
This is possible via the scaling theory for arbitrary polymer
networks derived by Duplantier. The obtained numbers do not vary much,
as due to the logarithmic dependence on the segment sizes. However,
they are all non-negligible, and therefore have to be compensated by
the released bond energy on formation of the DNA loop. We noted that
the entropy loss is of the same order or close to the bond melting
energy required for splitting an AT or GC bond, i.e., a considerable
amount. Moreover, it is to be expected that the vertex
effect increases the characteristic bond formation times in analytical
approaches which are based on the free energy.

Our calculations are valid in the long chain limit. In units of the
monomer size of a typical DNA double-helix persistence length
$\ell_p \sim 100$\,bp,
a minimum number of at least 10 monomers is expected to be
required to consider a segment in the final structure flexible. For
shorter segments, additional effects due to bending and twisting 
energy are expected to become relevant. As the mentioned examples
document, there are numerous systems, both in vivo and in vitro, in
which the flexibility conditions is easily fulfilled, and in which
our estimation method for the entropy loss becomes fully applicable.
The persistence length of single-stranded DNA and RNA is much shorter, 
typically taken to be of the order 
$\ell_p \sim 8$ bases. Thus, in single
strand looping experiments the expected entropy loss will be 
considerably larger.

\begin{acknowledgments}
We thank Sine Svenningsen for helpful discussions. This work was 
supported in part by the EPSRC (A.H.) and by the Emmy Noether 
programme of the Deutsche Forschungsgemeinschaft (R.M.).
\end{acknowledgments}

\begin{appendix}

\section{Configuration exponents for a general polymer network}

\begin{figure}
\unitlength=1cm
\begin{picture}(8,4.6)
\put(-2.4,-21.5){\includegraphics{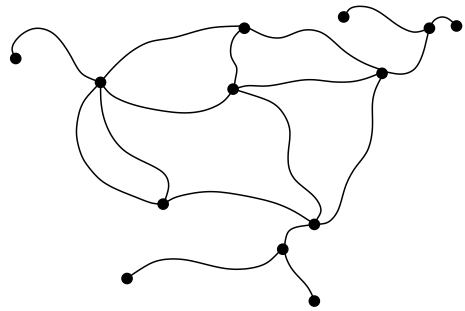}}
\end{picture}
\caption{Polymer network ${\cal G}$ with vertices ($\bullet$) of different
order ($n_1=5$, $n_3=4$, $n_4=3$, $n_5=1$).
\label{netw}}
\end{figure}
\begin{table}
\begin{tabular}{c|r|r|r}
\hline\hline
$N$ & $\sigma_N$ Pad{\'e} & $\sigma_N$ Pad{\'e}-Borel & $\sigma_N$ Pad{\'e}-Borel\\\hline
3 & -0.19 & -0.19 & -0.17\\
4 & -0.48 & -0.49 & -0.47\\
5 & -0.86 & -0.87 & -0.84\\
6 & -1.33 & -1.29 &      \\
7 & -1.88 & -1.75 &      \\
8 & -2.51 & -2.23 &      \\
9 & -3.21 & -2.71 &      \\\hline\hline
\end{tabular}
\caption{Topological exponents $\sigma_N$ for network vertices of order $N$
in 3D from various approximate techniques. The columns correspond to the first
three columns in Table 1 in
\protect\cite{schaefer}, where the scaling relation
$\gamma_F - 1 = \sigma_N + N \sigma_1$ (with $\gamma_F$ from \protect\cite{schaefer})
and $\sigma_1 = (\gamma-1)/2$ with the best
available value $\gamma = 1.1575$ \protect\cite{caracciolo}
have been used.
Note the large discrepancy between the different methods for increasing $N$.
\label{tab}}
\end{table}

A general polymer network ${\cal G}$ like the one depicted in Fig.~\ref{netw}
consists of vertices which are joined by ${\cal N}$ chain
segments of lengths $s_1,\ldots,s_{\cal N}$. Their total length be $L=\sum_{
i=1}^{\cal N} s_i$. In the scaling limit $s_i\gg 1$, the number of
configurations of such a network is given by
\cite{duplantier,schaefer,binder}
\begin{equation}
\label{network}
\omega_{\cal G} = A_{\cal G} \,
\mu^Ls_{\cal N}^{\gamma_{\cal G}-1}{\cal Y}_{\cal G}
\left(\frac{s_1}{s_{\cal N}},\ldots,\frac{s_{{\cal N}-1}}{s_{\cal N}}\right) \, ,
\end{equation}
where $A_{\cal G}$ is a non-universal amplitude, 
$\mu$ is the effective connectivity constant for self-avoiding walks,
and ${\cal Y}_{\cal G}$ is a scaling function. The topology of
the network is reflected in the configuration exponent
\begin{equation}
\gamma_{\cal G}=1-3\nu{\cal L}+\sum_{N\ge 1}n_N\sigma_N \, .
\end{equation}
${\cal L}=\sum_{N\ge 1}(N-2)n_N/2+1$ is the Euler number of independent 
loops, $n_N$ is the number of $N$-vertices, 
and $\sigma_N$ is an exponent connected to an $N$-vertex. Thus, expression
(\ref{network}) generalises the familiar form $\omega\sim\mu^LL^{\gamma-1}$
of a linear polymer chain. The numerical values we use in the text are
given in table \ref{tab} for the topological exponents $\sigma_N$;
furthermore, we employ $\nu=0.588$ and $\gamma\simeq1.16$ \cite{GZJ,caracciolo}. 
We also make use of the relation $\gamma
=2\sigma_1+1$. 

Note that in this work we consider the {\em entropy\/} of a given polymer 
network,
in which enters the total number of physically distinct configurations. 
Two configurations are considered distinct if they cannot be superimposed 
by translation. In particular, the monomers of the chain are distinguishable.
For a simple ring of length $L$ this implies that two configurations are
distinct even if they have the same trajectory, but differ from each other
by a reptation (translation of the chain within the trajectory)
by a non-integer multiple of $L$. The number of 
configurations of the simple ring is therefore \cite{duplantier}
\begin{equation}
\omega_{\rm circ} = \widetilde{\omega} L \sim L^{-3 \nu}
\end{equation}
where $\widetilde{\omega} \sim L^{- 3 \nu - 1}$ is the number of 
configurations of a ring polymer with indistinguishable monomers. Likewise, 
$\omega_{\rm circ}$ corresponds to the number of closed random walks 
of length $L$ which start and end at a given point in space (compare 
also the first reference \cite{slili2d}).

The number of configurations of a looped structure (with a least 
one vertex) is also given by equation (\ref{network}). This is due
to the fact that the established looping bond is chemically 
fixed within the chain, so that the chain cannot reptate within 
a given trajectory. For the same reason (and in contrast to 
references \cite{slili2d,hame}), different loops cannot exchange 
length with each other.

\end{appendix}


\begin{thebibliography}{99}
\bibitem{rem} It can be assumed that this criterion is fulfilled if the
segment contains more than ten persistence lengths.
\bibitem{remark} For a Gaussian random walk on a cubic lattice, 
$\ln (A_8 \, a /A_{\rm circ}) = \ln (A_8 \, {\cal Y}_8(1)/A_{\rm circ}) =
\ln [3/(2 \pi)] \simeq - 0.74$ is small compared with the
other contributions in equations (\ref{bond1}) and (\ref{half}).
\end{thebibliography}
\end{document}